\documentclass[runningheads]{llncs}
\usepackage[T1]{fontenc}
\usepackage{cite}
\usepackage{graphicx}
\usepackage{multirow}
\usepackage[dvipsnames]{xcolor}
\usepackage{hyperref}

\hypersetup{hypertex=true,
            colorlinks=true,
            linkcolor=blue,
            anchorcolor=blue,
            citecolor=blue}
\usepackage[misc]{ifsym}
\usepackage{fontawesome}
\usepackage{color}
\usepackage{amsfonts}
\newcommand\blfootnote[1]{%
  \begingroup
  \renewcommand\thefootnote{}\footnote{#1}%
  \addtocounter{footnote}{-1}%
  \endgroup
}

\begin{document}
\title{Interpretable Few-Shot Retinal Disease Diagnosis with Concept-Guided Prompting of Vision-Language Models}
\titlerunning{Concept-guided Prompting for Vision Language Models}
% If the paper title is too long for the running head, you can set
% an abbreviated paper title here
%

\author{Deval, Mehta\inst{1,2}\textsuperscript{*(\faEnvelopeO)}
\and Yiwen Jiang\inst{1,2}\textsuperscript{*} \and Catherine L Jan\inst{3,4} \and Mingguang He\inst{4,5,6,7} \and Kshitij Jadhav\inst{8} \and Zongyuan Ge\inst{1,2,9}}
%index{Mehta, Deval}
%index{Sivathamboo, Shobi}
%index{Simpson, Hugh}
%index{Kwan, Patrick}
%index{Terence, O'Brien}
%index{Ge, Zongyuan}
%
\authorrunning{Mehta, Jiang et al.}
% First names are abbreviated in the running head.
% If there are more than two authors, 'et al.' is used.
%
\institute{AIM for Health Lab, Faculty of IT, Monash University, Melbourne, Australia \and Faculty of Engineering, Monash University, Melbourne, Australia \\ \and 
Centre for Eye Research Australia, Royal Victorian Eye and Ear Hospital, East Melbourne, Victoria, Australia \\ \and
Ophthalmology, Department of Surgery , The University of Melbourne, Melbourne, Victoria, Australia \\ \and
Research Centre for SHARP Vision (RCSV), The Hong Kong Polytechnic University, Kowloon, Hong Kong \\ \and
Centre for Eye and Vision Research (CEVR), 17W Hong Kong Science Park, Hong Kong \\ \and
Department of Optometry and Vision Sciences, The University of Melbourne, Melbourne, Australia \\ \and
Koita Centre of Digital Health, Indian Institute of Technology Bombay, India \\ \and
Airdoc-Monash Research Lab, Monash University, Melbourne, Australia}
%Department of Neuroscience, Central Clinical School, Faculty of Medicine Nursing and Health Sciences, Monash University, Melbourne, Australia \\ \and
%Department of Neurology, Alfred Health, Melbourne, Australia \\ \and Departments of Medicine and Neurology, The University of Melbourne, Royal Melbourne Hospital, Parkville, Victoria, Australia \\ \and Airdoc-Monash Research Lab, Monash University, Melbourne, Australia}
%\institute{Princeton University, Princeton NJ 08544, USA \and
%Springer Heidelberg, Tiergartenstr. 17, 69121 Heidelberg, Germany
%\email{lncs@springer.com}\\
%\url{http://www.springer.com/gp/computer-science/lncs} \and
%ABC Institute, Rupert-Karls-University Heidelberg, %Heidelberg, Germany\\
%\email{\{abc,lncs\}@uni-heidelberg.de}}
%
\maketitle              % typeset the header of the contribution
\begin{abstract}
Recent advancements in deep learning have shown significant potential for classifying retinal diseases using color fundus images. However, existing works predominantly rely exclusively on image data, lack interpretability in their diagnostic decisions,  and treat medical professionals primarily as annotators for ground truth labeling. To fill this gap, we implement two key strategies: extracting interpretable concepts of retinal diseases using the knowledge base of GPT models and incorporating these concepts as a language component in prompt-learning to train vision-language (VL) models with both fundus images and their associated concepts. Our method not only improves retinal disease classification but also enriches few-shot and zero-shot detection (novel disease detection), while offering the added benefit of concept-based model interpretability. Our extensive evaluation across two diverse retinal fundus image datasets illustrates substantial performance gains in VL-model based few-shot methodologies through our concept integration approach, demonstrating an average improvement of approximately \textbf{5.8\%} and \textbf{2.7\%} mean average precision for 16-shot learning and zero-shot (novel class) detection respectively. Our method marks a pivotal step towards interpretable and efficient retinal disease recognition for real-world clinical applications.\blfootnote{* Authors contributed equally}

\end{abstract}
\section{Introduction}
Globally, retinal disorders are significant contributors to ocular morbidity and visual impairment~\cite{thapa2020prevalence}. Retinal diseases like age-related macular degeneration (AMD) and diabetic retinopathy (DR) are the leading cause of irreversible blindness in older populations and working-age individuals in developed countries ~\cite{kempen2004prevalence, friedman2004prevalence}. Notably, the initial stages of AMD and DR often present with only subtle symptoms but can progress to severe, irreversible optic nerve damage if not addressed promptly~\cite{kocur2002visual}. Thus, the importance of early detection and accurate diagnosis of retinal diseases is crucial in averting permanent alterations to vision. 

Recent advancements in diagnostic technologies, including Fundus photography (for retina, optic disc and macula), Optical Coherence Tomography (OCT), and Fundus Fluorescein Angiography (FFA), have significantly increased the accessibility of routine eye disease screenings within healthcare and ophthalmology practices, thereby enhancing the potential for early intervention~\cite{panwar2016fundus}. Of these techniques, color fundus imaging (illumination of retina by white light) remains the most practical and straightforward method for ocular examination due to its efficiency and simplicity. Especially, the advent of compact, portable cameras has bolstered the utility of color fundus imaging in teleophthalmology applications, particularly in  remote areas~\cite{bahl2022diabetic}. Nonetheless, the manual analysis of fundus images demands considerable expertise and time from ophthalmologists, a challenge that has been exacerbated during the recent public health crisis~\cite{nair2020effect}.

Several studies have demonstrated the efficacy of utilizing deep learning (DL) techniques for automated analysis of color fundus images, successfully identifying prevalent retinal diseases such as DR, glaucoma, and AMD~\cite{fu2018disc,araujo2020dr,li2021applications}. However, detecting rare eye diseases is a challenge due to data scarcity. Recent DL techniques have extended to the diagnosis of less common disorders, including Coat's disease and retinoschisis~\cite{gao2023discriminative,ju2021relational}. There have also been few works which concentrate on developing few-shot capability (i.e., using only a few examples of a disease category during training) of DL models to detect both common and rare retinal diseases~\cite{quellec2020automatic,murugappan2022novel,chen2023dynamic}. Developing few-shot capability is vital for diagnosing rare diseases where sample scarcity hinders effective training of DL models. 

Authors in ~\cite{quellec2020automatic} developed an unsupervised probabilistic model to detect rare eye diseases based on the latent feature embeddings of a trained Convolutional Neural Network (CNN) on common eye diseases, whereas~\cite{murugappan2022novel} adapted prototype networks for episodic learning to learn an enhanced latent space representation for few-shot detection. Beyond enhancing feature space representation, certain studies have focused on leveraging knowledge distillation across various eye diseases and implementing distinct sampling approaches to detect rare diseases and cultivate few-shot learning capabilities. For example, in the relational subset knowledge distillation~\cite{ju2021relational} divided the long-tailed data into multiple subsets based on the region and phenotype features of eye diseases and employed knowledge distillation across these subsets. Similarly, the hierarchical knowledge guided learning~\cite{ju2023hierarchical} demonstrated that prior relational knowledge between different ophthalmic diseases represented in a hierarchical manner along with instance-wise class balanced sampling could enhance rare disease detection.

However, despite advancements in few-shot learning using color fundus images, existing approaches face two limitations - 1) They rely solely on color fundus imagery, employing strategies like knowledge transfer, subsampling, or enhancing latent space disease feature representation. 2) They lack interpretability, relying primarily on GradCAM-based attention map explanations~\cite{selvaraju2017grad} and employ domain knowledge only limited to ground truth disease labels. These approaches overlooks the rich domain knowledge that could inform model development, especially the explicit consideration of disease-specific concepts and descriptive attributes, commonly used by experts to reach a diagnosis.

\begin{figure}[t!]
\centering
\includegraphics[width=0.95\textwidth]{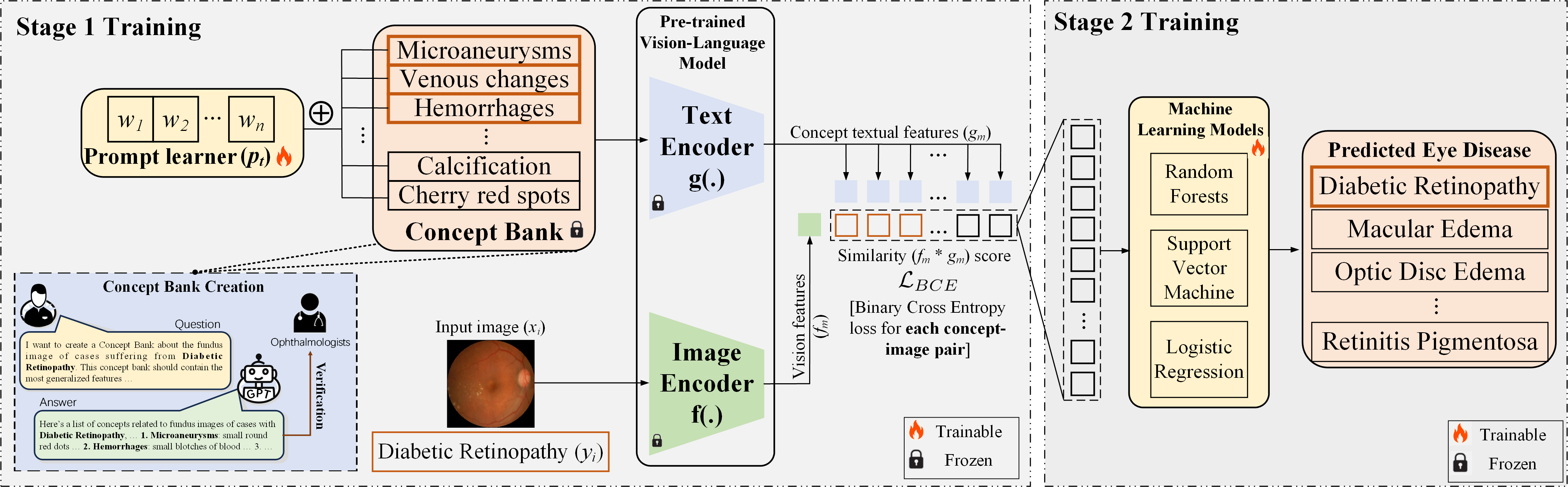}
\caption{Our framework consists of two stages: \textbf{Stage 1} focuses on developing a concept bank through GPT and validation by ophthalmologists, followed by concept-guided prompt learning of VL models for concept prediction. \textbf{Stage 2} trains machine learning models to classify disease categories based on Stage 1's concepts.} \label{fig1}
\end{figure}

\noindent\textbf{Our contributions}: In this work, we tackle the above two challenges in two steps: first, by constructing a concept bank that encapsulates domain knowledge specific to retinal eye diseases from color fundus images; and second, by incorporating this knowledge into the training of vision-language (VL) models, such as CLIP~\cite{radford2021learning}. Constructing concept banks manually is a labor-intensive and resource-demanding process that requires highly specialized expertise. To address this critical limitation, our approach leverages the extensive knowledge base of large language models, such as GPT-3~\cite{brown2020language}, to generate a concept bank for retinal eye diseases. This automated process significantly reduces the dependency on resource-intensive manual efforts while capturing diverse detailed medical knowledge. To ensure reliability and eliminate inaccuracies, our generated concept bank is validated by expert ophthalmologists, transforming their role from exhaustive creation to efficient verification. This streamlined process not only enhances accuracy but also improves scalability, making it an efficient approach for integrating accurate domain knowledge into medical AI systems.

Subsequently, we incorporate this concept bank into the model training process by leveraging a prompt-tuning paradigm to integrate domain knowledge with VL models. Prompt learning, a recent trend in Natural Language Processing (NLP), has emerged as an efficient approach for adapting VL models to downstream tasks. Recently, prompt learning techniques such as Context Optimization (CoOp)~\cite{zhou2022conditional} and Conditional Context Optimization (CoCoOp)~\cite{zhou2022learning} have demonstrated their effectiveness in classification tasks, particularly in few-shot learning scenarios. Despite their efficiency these methods lack interpretability. Thus, building on these advancements, we adapt prompt learning techniques and integrate our concept bank to develop an interpretable concept-guided prompt-tuning paradigm for VL models. This approach empowers our framework to predict both disease-specific concepts and categories effectively. 

We evaluate our framework on two color fundus imaging datasets - an in-house dataset comprising of 29 retinal diseases and the public RFMiD~\cite{pachade2021retinal} dataset. Our evaluation results demonstrate significant mean average precision (mAP) improvements in both few-shot and zero-shot detection capability of VL models by ~\textbf{5.8\%} and ~\textbf{2.7\%} respectively. Importantly, our proposed Concept-Guided Prompting framework for Vision-Language Models enhances interpretability, even for previously unseen retinal disease categories, thereby increasing the trustworthiness and reliability of automated diagnostic systems.

\section{Proposed Method}
We present our overall framework in Fig~\ref{fig1}, consisting of two training stages. The first stage involves predicting concepts within the input image, while the second stage focuses on determining the disease category from these predicted concepts.

\subsection{Stage 1: Concept-guided Prompting of Vision-Language Models}
We first describe the process of building the concept bank, which is necessary for concept-guided training.

\subsubsection{Accurate Construction of Retinal Diseases Concept Bank:}
Large Language Models (LLMs) like GPT-4, trained on substantial medical knowledge, have achieved scores comparable to passing the United States Medical Licensing Examination~\cite{gilson2023does} and have been shown to produce concept banks with greater factuality and groundability than human-written content in specific domains~\cite{yang2023language}. We leverage the embedded knowledge of LLMs to simplify the creation of our retinal disease concept bank, reducing dependence on specialists, as shown in Fig~\ref{fig1}.

\begin{figure}[t!]
\centering
\includegraphics[width=0.85\textwidth]{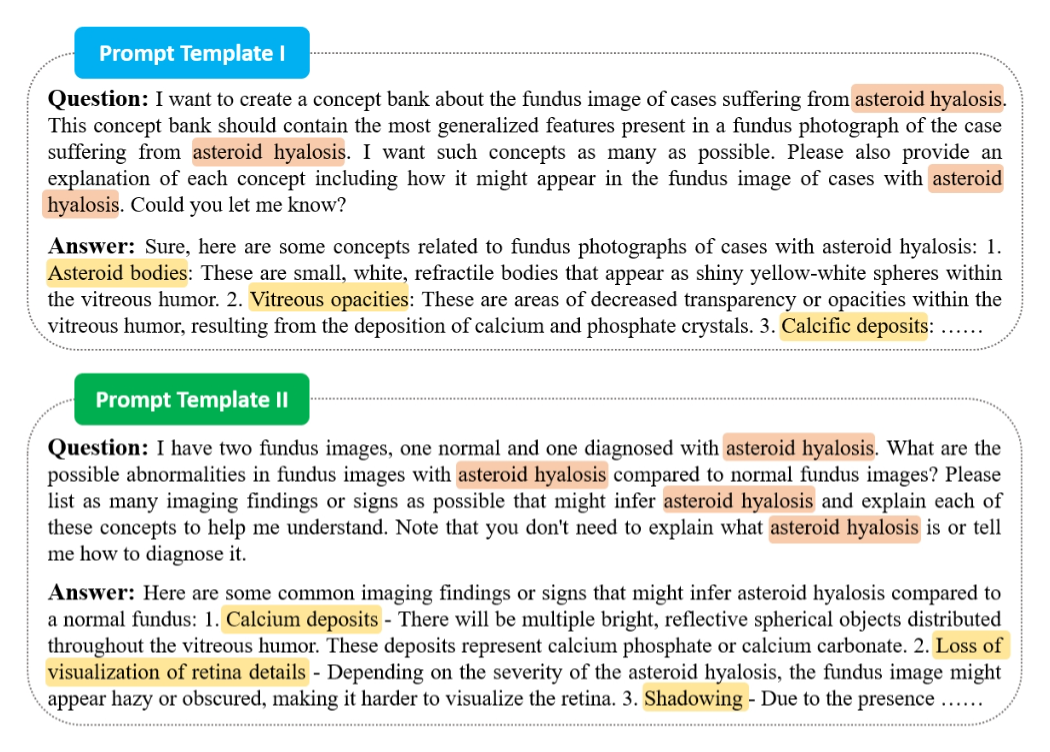}
\caption{Design of two prompt templates for constructing the retinal disease concept bank. The figure illustrates the prompts for the retinal disease of Asteroid Hyalosis with adaptable disease-specific parts. These two templates prompt LLMs twice, generating concepts that are later validated by ophthalmologists.} \label{fig2}
\end{figure}

Figure~\ref{fig2} illustrates our two adaptable prompting templates for extracting key concepts related to retinal diseases. Template 1 explicitly requests concepts for a specified condition, while Template 2 generates diagnostic concepts by comparing the condition to a normal retinal fundus image. To ensure consistency, each prompt is used twice, and only overlapping or common concepts across all prompts are retained. This process is applied across all retinal conditions in our dataset to construct a comprehensive concept bank. 

To mitigate risks of misinformation~\cite{kaddour2023challenges} and ensure accuracy, the generated concepts are validated by ophthalmologists. For example, as shown in Fig~\ref{fig2}, concepts for Asteroid Hyalosis generated by Template 1 include asteroid bodies, vitreous opacities, and calcific deposits, while Template 2 identifies calcium deposits, loss of retinal detail visualization, and shadowing. Overlapping concepts such as calcium deposits, vitreous opacities, and asteroid bodies were selected as representative and confirmed by ophthalmologists as attributes arising from calcium and phosphate crystal deposits commonly associated with Asteroid Hyalosis~\cite{bergren1991prevalence,tripathy2018asteroid}.

\textbf{Our in-house dataset includes 77 concepts spanning 29 retinal diseases, while the RFMiD dataset contains 119 concepts across 46 retinal diseases, with several categories sharing common concepts}. This streamlined and validated approach ensures high accuracy, efficiency, and reliability in generating concept banks for medical AI applications.

\subsubsection{Vision Language (VL) Models:}
We commence with an overview of VL models, specifically highlighting CLIP~\cite{radford2021learning}. CLIP features two distinct encoders: an image encoder and a text encoder, as illustrated in Fig~\ref{fig1}. The image encoder, typically based on Resnet50~\cite{he2016deep} or ViT~\cite{dosovitskiy2020image}, learns a low-dimensional feature representation of the input image. Conversely, the text encoder, built on the Transformer architecture~\cite{vaswani2017attention}, processes natural language input (usually "a photo of [CATEGORY NAME]") to generate text representations. CLIP is then trained to align these two embedding spaces by enhancing the cosine similarity for matching image-text pairs and reducing it for non-matching ones. Importantly, CLIP has been trained on a vast dataset of 400 million image-text pairs, enabling it to capture a wide range of visual concepts transferable to downstream tasks. Thus, we utilize the pre-trained CLIP encoders for their broad feature representation capabilities, keeping them frozen during our model training as depicted in Fig~\ref{fig1}.

\subsubsection{Integrating concepts with VL Models:}
Let $X=\{x_{i}\}_{i=1}^{N}$ represent the set of $N$ input images. Each image $x_i$ belongs to a ground truth disease category $y_i \in \left\{1, 2, \ldots, K\right\}$ of the total $K$ categories and is also associated with a subset of concepts $C_i$ from the concept bank $C=\{c_1, c_2, \ldots, c_E\}$, containing $E$ concepts. In Fig~\ref{fig1}, the input image $x_i$'s concepts and disease category are marked in brown. Then, the few-shot learning task involves classifying $x_i$ correctly with minimal training examples of $y_i$.

Drawing on recent advancements in prompt learning techniques such as CoOp~\cite{zhou2022conditional} and CoCoOp~\cite{zhou2022learning}, which enhance few-shot detection in vision-language models, we devise a set of learnable prompt vectors ($p_t$). In our approach, this prompt is  uniquely terminated (see ablation study for different termination token positions) with the concept name, given by below.

\begin{equation}\label{eq1}
p_t = [w_1] [w_2] ... [w_M] [CONCEPT] 
\end{equation}

where $[CONCEPT]$ represents a single concept from the concept bank set $C$, and each $[w_i] \in \{\ 1, 2, ... M \} $ is a vector with the same dimension as word embeddings (i.e., 512 for CLIP) and $M$ is a hyperparameter specifying the number of prompt tokens. Each of the 512 vector prompt tokens $[w_1,..,w_M]$ are randomly initialized and represent words in CLIP's 49,152-size vocabulary. These tokens are shared across all concepts. \textbf{The prompt tokens and ML models are the only trainable components as marked in Fig~\ref{fig1}}.

The prompt $p_t$ is then given to CLIP's frozen text encoder g($\cdot$), generating textual concept features $g_m = g(p_t)$. Concurrently, the input image $x_i$ is processed by CLIP's frozen image encoder to yield the visual features $f_m = f(x_i)$.

After extracting textual concept features and visual features, similarity scores ($f_m * g_m$) are calculated, as depicted in Fig~\ref{fig1}. Given an image contains multiple concepts, its visual features must align with the textual features of these concepts, framing this as a multi-label classification problem. Therefore, the concepts are passed one-by-one as a CONCEPT TOKEN in Eq ~\ref{eq1} and then disease-specific concepts are matched via BCE loss as highlighted in Fig~\ref{fig1} and given by below.

\begin{equation}\label{eq2}
Concept_{loss} = -\frac{1}{E} \sum_{j=1}^{E} \left[ c_j \log(\hat{c}_j) + (1 - c_j) \log(1 - \hat{c}_j) \right]
\end{equation}

where $E$ represent the total number of concepts, $c_j$ indicates the ground truth presence (1 for present, 0 for absent) of the $j^{th}$ concept in image $x_i$, $\hat{c}_j$ denotes the predicted probability for the $j^{th}$ concept, and $Concept_{loss}$ is the mean BCE loss across all concepts.

In Fig~\ref{fig1}, the input image classified as Diabetic Retinopathy (DR), features key concepts like Microaneurysms, Hard exudates, and Hemorrhages, highlighted in brown. Our optimization aims to predict these concepts accurately, using BCE loss to identify these three concepts (as 1) and exclude others (as 0). For images with multiple diseases, all pertinent concepts are targeted for prediction. The optimization goal is to minimize $Concept_{loss}$, with gradients backpropagated through the text encoder $g(\cdot)$, thereby training the prompt learner parameters via matching concept features between the image and text.

\subsection{Stage 2: Predicting Disease Category from Concepts}

Following stage 1's concept prediction training, we utilize concept logits for disease categorization using the concept-bottleneck model (CBM)\cite{koh2020concept}. The key idea behind CBM is the introduction of an intermediate concept bottleneck layer before the final classification stage, allowing the model to predict interpretable concepts alongside the final categories. This design significantly enhances the interpretability of deep learning models by providing insights into how specific concepts contribute to the predictions. For mapping concept logits to disease categories, we employ machine learning techniques such as Random Forest (RF), Logistic Regression (LR), and Support Vector Machine (SVM). Although end-to-end mapping with a Multi-layer Perceptron (MLP), as suggested by CBM\cite{koh2020concept}, is an alternative, our experiments demonstrate that traditional ML models outperform MLPs for our dataset, as discussed in the ablation study later. We adhere to standard training procedures for RF, LR, and SVM, detailed below.

\begin{equation}\label{eq3}
\hat{y}_i = \frac{1}{K} \sum_{k=1}^{K} f_{i, k}(c_x)
\end{equation}

\begin{equation}\label{eq4}
\hat{y}_i = \frac{1}{1 + e^{-(w_i \cdot c_x + b_i)}}
\end{equation}

\begin{equation}\label{eq5}
\hat{y}_i = sign(w_i \cdot c_x + b_i)
\end{equation}

where $\hat{y}_i$ denotes the predicted probability for the $i^{th}$ disease category, with $c_x$ being the input logits for image $x_i$'s predicted concepts from stage 1. For LR and SVM, $w_i$ and $b_i$ represent the weight and bias, respectively, whereas, $f_{i,k}$ is the $k^{th}$ decision tree's prediction in the RF model.

\subsection{Training \& Implementation Details}
To construct the initial concept bank, we utilized the GPT-3.5-Turbo Model API. Our method was trained on the ViT-B16 CLIP pre-trained model, incorporating standard data augmentations—random flip, center crop, and horizontal translation—and resizing input images to 224x224. We started with a learning rate of $1 \times 10^{-3}$, applying a cosine learning rate scheduler, a warm-up phase of 5 epochs, and a constant schedule type. Training utilized a batch size of 32, SGD optimizer, and spanned 100 epochs across both in-house and public datasets. Following existing long-tailed works~\cite{ju2023hierarchical, ju2021relational}, the evaluation metric was mean average precision (mAP), with few-shot and zero-shot detection experiments averaged over five runs with different random seeds for training shots. Experiments were conducted on PyTorch with an NVIDIA A100 GPU.

\section{Datasets}
\subsubsection{In-house dataset:}
Our in-house dataset comprises 4,535 color fundus images (captured with Topcon TRC-NW8 camera) across 29 disease categories, annotated for ground truth retinal disease by two senior ophthalmologists. This long-tailed dataset predominantly features diseases like Macular edema, Central Retinal vein occlusion, and Diabetic Retinopathy, with fewer instances of rarer conditions like Anterior ischemic optic neuropathy, Choroiditis, Choroidal hemangioma, and Macular dystrophy. The dataset is divided into training, validation, and test sets, containing 2928, 727, and 880 images respectively.

\subsubsection{RFMiD} - The RFMiD dataset~\cite{pachade2021retinal} contains 3,200 fundus images across 46 disease categories, annotated by two senior retinal experts. This long-tailed dataset primarily features common diseases like Diabetic Retinopathy, Media Haze, Optic Disc Cupping, but includes scarce instances of rare diseases such as Coloboma and Hemorrhagic Retinopathy. The dataset is pre-divided into training, validation, and testing sets with 1920, 640, and 640 images respectively. It's a multi-label dataset, with 749 images bearing multiple labels and 2,451 images with a single label. Our approach treats this dataset with a multi-label classification loss function.

\section{Results}
We evaluate our proposed framework in both few-shot and zero-shot settings, benchmarking its performance against state-of-the-art prompt learning approaches - CoOp~\cite{zhou2022learning}, CoCoOp~\cite{zhou2022conditional}, ProDA~\cite{lu2022prompt}, MaPLe~\cite{khattak2023maple} for vision-language (VL) models. Additionally, we conduct interpretability analyses to assess the framework's effectiveness and perform comprehensive ablation studies to identify the optimal hyperparameters for our approach.

\subsection{Few-shot Classification}
In few-shot classification, we randomly select $n$ images per category from the training fold and test it on all the images from the test fold. Table~\ref{tab1} shows that image-only baseline performs the poorest, with improvements seen in fine-tuned CLIP, further enhanced by prompt learning, and maximized with our concept-guided approach. RFMiD dataset results are lower than our in-house dataset due to its multi-label classification complexity. Notably, our MaPLe + Concept method outperforms others in both datasets, benefiting from multi-modal prompts that bolster vision-language generalization. The average increment of all our concept-guided prompt learning methods increases is \textbf{6.45\%} and \textbf{5.15\%} for 16-shot detection for our In-house and RFMiD dataset respectively. 

\begin{table}[t!]
\caption{Few-shot classification enhancement for different $n$ shots (in terms of $mean_{std}$ mAP of 5 random seeds) with concept-guided training across prompt learning methods on in-house and public datasets (best viewed in zoom).} \label{tab1}
\resizebox{\textwidth}{!}{\begin{tabular}{c|c|c|c|c|c|c|c|c}
\hline
\multirow{2}{*}{Method/ Dataset / Shots}&\multicolumn{4}{c|}{In-house dataset} &\multicolumn{4}{|c}{RFMiD}\\
\cline{2-9}
 & n=2 & n=4 & n=8 & n=16 & n=2 & n=4 & n=8 & n=16 \\
\hline
Image-only (ViTB-16) & $29.36_{1.68}$ & $32.41_{1.97}$ & $35.67_{1.62}$ & $38.71_{1.43}$ & $27.58_{1.85}$ & $30.46_{1.37}$ & $33.59_{1.5}$ & $35.64_{1.26}$ \\
CLIP~\cite{radford2021learning} (linear probing) & $36.75_{2.07}$ & $41.84_{2.23}$ & $44.73_{1.93}$ & $48.38_{0.98}$ & $34.27_{2.19}$ & $38.42_{1.95}$ & $41.75_{2.1}$ & $46.81_{1.22}$ \\
\hline
CoOp~\cite{zhou2022learning} & $39.43_{1.32}$ & $47.71_{1.5}$ & $51.52_{2.05}$ & $55.64_{1.45}$ & $37.89_{2.06}$ & $40.63_{1.72}$ & $48.93_{1.49}$ & $52.41_{1.45}$  \\
CoCoOp~\cite{zhou2022conditional} & $39.36_{1.38}$ & $47.49_{1.57}$ & $51.06_{1.56}$ & $53.97_{1.20}$ & $37.75_{2.12}$ & $40.44_{1.89}$ & $47.61_{1.39}$ & $51.42_{1.40}$  \\
ProDA~\cite{lu2022prompt} & $39.72_{1.53}$ & $47.24_{1.87}$ & $51.31_{1.47}$ & $55.02_{1.33}$ & $37.61_{1.56}$ & $41.43_{1.51}$ & $48.79_{1.55}$ & $53.18_{1.21}$ \\
MaPLe~\cite{khattak2023maple} & $57.64_{2.28}$ & $58.36_{1.02}$ & $62.58_{1.45}$ & $66.32_{1.27}$ & $49.44_{1.76}$ & $51.84_{1.89}$ & $58.62_{1.51}$ & $61.53_{1.65}$  \\
\hline
CoOp + Concepts (Ours) & $40.9_{2.83}$ & $51.43_{1.32}$ & $55.8_{1.46}$ & $61.67_{0.33}$ & $39.27_{1.73}$ & $44.83_{2.24}$ & $52.25_{1.64}$ & $57.47_{1.41}$  \\
CoCoOp + Concepts (Ours) & $40.61_{2.28}$ & $50.94_{1.5}$ & $55.39_{1.68}$ & $61.26_{0.59}$ & $38.96_{1.37}$ & $45.58_{1.30}$ & $51.46_{1.21}$ & $57.26_{1.32}$ \\
ProDA + Concepts (Ours) & $42.16_{1.61}$ & $52.08_{1.51}$ & $56.26_{1.98}$ & $63.14_{1.25}$ & $41.91_{1.66}$ & $45.59_{1.38}$ & $54.62_{1.53}$ & $59.35_{1.10}$  \\
MaPLe + Concepts (Ours) & $\textbf{59.47}_{1.55}$ & $\textbf{61.9}_{0.88}$ & $\textbf{65.80}_{1.8}$ & $\textbf{70.67}_{1.32}$ & $\textbf{50.57}_{1.49}$ & $\textbf{55.4}_{2.37}$ & $\textbf{60.57}_{1.96}$ & $\textbf{65.07}_{1.94}$  \\
\hline
\hline
Average Increment & $1.75$ & $3.89$ & $4.19$ & $6.45$ & $2.01$ & $4.26$ & $3.74$ & $5.15$ \\
\hline
\end{tabular}}
\end{table}

\subsection{Base to Novel Generalization (Zero-shot Detection)}
Our framework's zero-shot detection capability is evaluated through base-to-novel generalization, training on common base categories and testing on unseen novel categories, which are underrepresented in the dataset. Models were trained with 16 shots from base categories, defined as the most common first half in each dataset, while novel categories were excluded from training. This resulted in 15 base and 14 novel categories for our in-house dataset, and 23 each for RFMiD.

\begin{table}
\centering
\caption{Zero-shot detection enhancement via concept-guided training in prompt learning methods on in-house and public datasets. Performance presented as $mean_{std}$.} \label{tab2}
\resizebox{0.85\textwidth}{!}{\begin{tabular}{c|c|c|c|c}
\hline
\multirow{2}{*}{Method/ Dataset}&\multicolumn{2}{c|}{In-house dataset} &\multicolumn{2}{|c}{RFMiD}\\
\cline{2-5}
 & weighted f1 & mAP & weighted f1 & mAP \\
\hline
Image-only (ViTB-16) & $15.63_{1.23}$ & $18.47_{0.66}$ & $14.82_{0.46}$ & $16.79_{0.51}$  \\
CLIP~\cite{radford2021learning} (linear probing) & $19.46_{0.65}$ & $21.52_{1.07}$ & $16.48_{1.29}$ & $18.33_{1.38}$  \\
\hline
CoOp~\cite{zhou2022learning} & $21.19_{0.65}$ & $23.85_{0.98}$ & $18.96_{1.18}$ & $19.42_{1.04}$  \\
CoCoOp~\cite{zhou2022conditional} & $25.82_{0.54}$ & $26.95_{1.19}$ & $19.71_{0.67}$ & $21.36_{0.54}$  \\
ProDA~\cite{lu2022prompt} & $25.16_{1.27}$ & $25.73_{0.55}$ & $19.64_{0.77}$ & $20.92_{0.70}$ \\
MaPLe~\cite{khattak2023maple} & $26.71_{1.19}$ & $28.38_{0.62}$ & $20.49_{1.59}$ & $21.61_{1.66}$  \\
\hline
CoOp + Concepts (Ours) & $24.31_{0.96}$ & $25.29_{0.98}$ & $20.74_{1.5}$ & $22.63_{1.18}$ \\
CoCoOp + Concepts (Ours) & $27.93_{0.52}$ & $29.41_{1.09}$ & $21.87_{0.84}$ & $23.5_{0.99}$ \\
ProDA + Concepts (Ours) & $27.29_{1.19}$ & $28.17_{1.26}$ & $21.35_{0.8}$ & $23.86_{0.53}$  \\
MaPLe + Concepts (Ours) & $\textbf{30.61}_{1.06}$ & $\textbf{32.8}_{0.8}$ & $\textbf{22.39}_{0.77}$ & $\textbf{24.17}_{0.66}$  \\
\hline
\hline
Average Increment & $2.81$ & $2.69$ & $1.89$ & $2.71$ \\
\hline
\end{tabular}}
\end{table}

The entire concept bank, representing knowledge of all diseases, was included during training, but no novel category samples were used. Testing focused solely on novel categories, making the task highly challenging due to the long-tailed nature of the datasets and the use of limited base samples, simulating real-world clinical conditions without fine-tuning on novel samples.

As shown in Table~\ref{tab2}, our concept-based approach consistently improves prompt learning methods, achieving an average mAP increase of \textbf{2.69\%} and \textbf{2.71\%} for the in-house and RFMiD datasets, respectively.

\subsection{Interpretability: From Concepts to Diseases}

\begin{figure}[t!]
\centering
\includegraphics[width=0.95\textwidth]{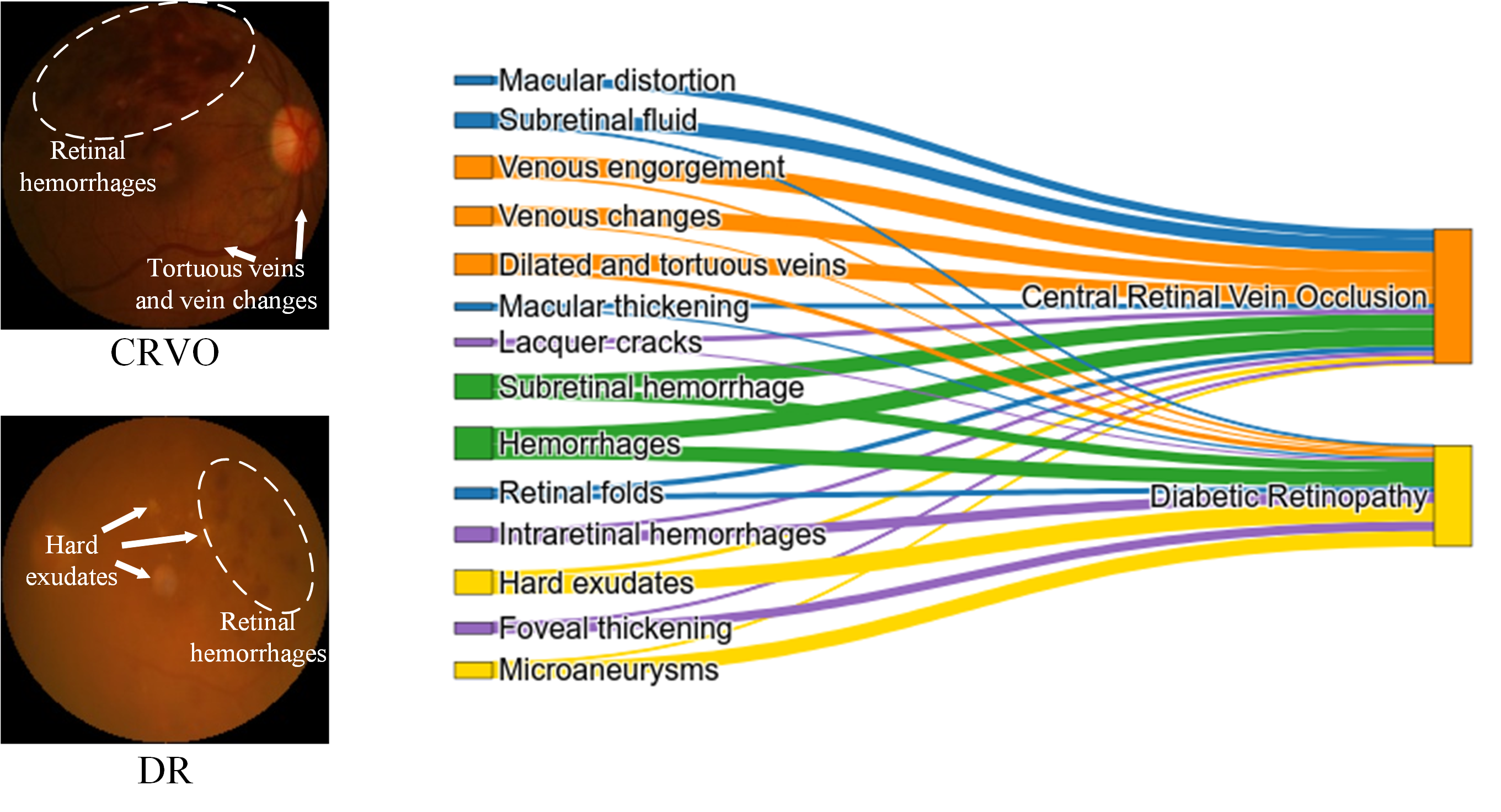}
\caption{Fundus images illustrating key representative attributes of Diabetic Retinopathy (DR) and Central Retinal Vein Occlusion (CRVO) and a sankey diagram depicting the flow of the concepts relevant to DR and CRVO learnt by our framework. We present the top 5 and the bottom 5 concepts associated with each DR and CRVO based on the contribution scores of the concepts to DR and CRVO from our LR model. The scores are averaged and normalized over all the test samples of DR and CRVO.} \label{fig3}
\end{figure}

\begin{figure}[t!]
\centering
\includegraphics[width=0.95\textwidth]{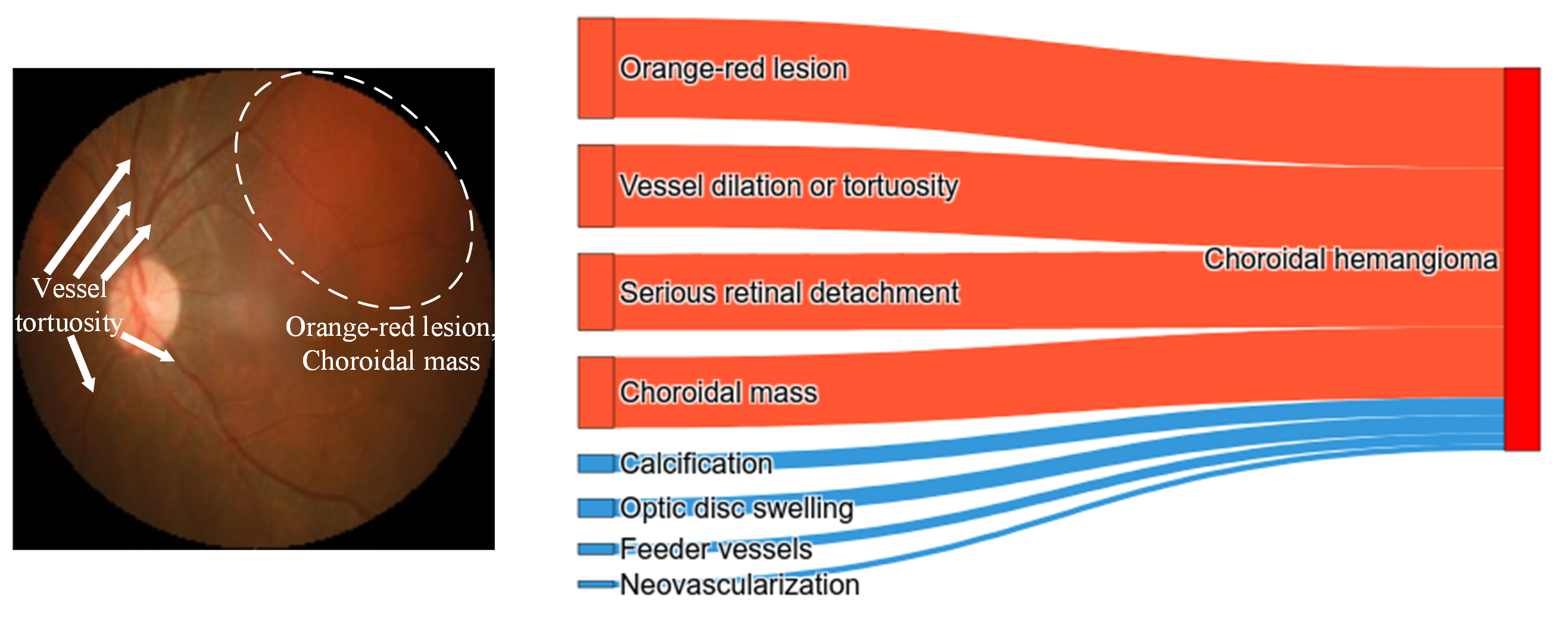}
\caption{Fundus image illustrating key representative attributes of Choroidal hemangioma and a sankey diagram depicting the flow of the concepts relevant to the novel category of Choroidal hemangioma in zero-shot detection setting of our framework. We present the top 4 and the bottom 4 concepts associated with choroidal hemangioma learnt by our model. The scores are averaged and normalized over all the test samples choroidal hemangioma.} \label{fig4}
\end{figure}

Our stage 2 training using Logistic Regression provides clarity on concept's contribution to the predicted disease, thus enhancing the interpretability of diagnosis. Concept representation enhances interpretability of predicted categories demonstrated in two scenarios: distinguishing similar categories in a few-shot setting (Fig~\ref{fig3}) and detecting novel categories in a zero-shot setting (Fig~\ref{fig4}).

\noindent\textbf{Few-shot Distinction:} We differentiate Diabetic Retinopathy (DR) and Central Retinal Vein Occlusion (CRVO), two retinal vascular diseases~\cite{park2016cell} which exhibit overlapping features like hemorrhages and subretinal hemorrhage~\cite{sen2021differences}. DR-specific features include hard exudates and microaneurysms~\cite{yau2012global}, while CRVO is characterized by venous changes and engorgements~\cite{hayreh2015fundus}. Fig~\ref{fig3} highlights these distinctions, where concept weights, averaged and normalized across test samples of DR and CRVO, are represented by the width of the flow and boxes. Shared concepts, such as hemorrhages, exhibit strong associations with both diseases, while distinguishing concepts (e.g., venous changes for CRVO, hard exudates for DR) have high associations with their respective conditions and minimal flow to the other. Unrelated concepts show negligible association with either condition.

\noindent\textbf{Zero-shot Detection:} In Fig~\ref{fig4}, we present the scenario for a rare condition choroidal hemangioma~\cite{shields1988choroidal}. We selected the top 4 most relevant and the bottom 4 least relevant concepts according to their contribution averaged over all the test samples of choroidal hemangioma. Our framework accurately identifies key concepts, including orange-red lesion~\cite{witschel1976hemangioma}, vessel dilation~\cite{lupidi2024new}, serous retinal detachment, and choroidal mass, with strong associations depicted by wide flows. Less relevant concepts, such as calcification, feeder vessels, and optic disc swelling~\cite{shanmugam2015vascular}, are correctly identified with minimal association, demonstrating the framework's ability to generalize and handle unseen categories effectively.

\subsection{Ablation study for concept token position and number of learnable tokens}

Previous studies~\cite{zhou2022learning} have shown that the position of the class token in prompt learning significantly impacts performance. Building on this insight, we investigate how the position of the [CONCEPT] token affects the performance of our retinal disease diagnosis framework. Specifically, we experiment with three token positions:  START $\{0\}$, MIDDLE $\{ceil(M/2)\}$, and END $\{M-1\}$, where $M$ represents the number of learnable prompt tokens. The performance variations for these configurations are presented in Table~\ref{concept_token}, with results showing that placing the [CONCEPT] token at the END position yields the best performance. This improvement may stem from the model leveraging the summary of all preceding learnable tokens at the END position, optimizing its ability to represent the concept in a holistic and comprehensive manner.

\begin{table}
\centering
\caption{Ablation study on [CONCEPT] position in our prompt learner, showing average mAP from 5 random seeds for 16-shot classification on our In-house dataset.} \label{concept_token}
\resizebox{0.65\textwidth}{!}{\begin{tabular}{c|c|c|c}
\hline
Method/ Position of Token & START & MIDDLE & END \\
\hline
CoOp + Concepts (Ours) & 59.91 & 58.35 & \textbf{61.67} \\
CoCoOp + Concepts (Ours) & 58.35 & 61.08 & \textbf{61.26} \\
ProDA + Concepts (Ours) & 62.93 & 63.07 & \textbf{63.14}  \\
MaPLe + Concepts (Ours) & 69.46 & 68.31 & \textbf{70.67}  \\
\hline
\end{tabular}}
\end{table}

We perform an additional ablation study for evaluating the performance variation with the number of learnable prompt tokens. The results of this ablation study are shown in Table~\ref{number_of_tokens}. We varied the number of tokens from 2 to 64 in powers of 2. We noticed that the performance increased as the number of tokens increased, and it saturated when the number of tokens was 32.

\begin{table}
\centering
\caption{Ablation study of number of learnable tokens in our prompt learner, showing average mAP from 5 random seeds for 16-shot classification on In-house dataset.} \label{number_of_tokens}
\resizebox{0.85\textwidth}{!}{\begin{tabular}{c|c|c|c|c|c|c}
\hline
Method/ Number of learnable tokens & 2 & 4 & 8 & 16 & 32 & 64\\
\hline
CoOp + Concepts (Ours) & 58.36 & 59.47 & 60.58 & 60.91 & \textbf{61.67} & 61.05\\
CoCoOp + Concepts (Ours) & 57.24 & 58.39 & 60.27 & 60.92 & \textbf{61.26} & 60.53\\
ProDA + Concepts (Ours) & 59.97 & 60.65 & 62.41 & 63.06 & \textbf{63.14} & 62.93\\
MaPLe + Concepts (Ours) & 67.94 & 68.33 & 70.04 & 70.18 & \textbf{70.67} & 70.24\\
\hline
\end{tabular}}
\end{table}

\subsection{Ablation study for stage 2 training strategy}

Table~\ref{tab5} presents our exploration of various machine learning methods for the stage 2 training strategy. We compared Logistic Regression (LR), Random Forest (RF), Support Vector Machine (SVM), and an end-to-end training strategy for n-shot classification. The results demonstrate that Logistic Regression consistently outperformed the other methods across all n-shot settings.

\begin{table}
\centering
\caption{Evaluation of Stage 2 models using Stage 1 MaPLe + Concepts, showing average mAP from 5 random seeds for n-shot classification on In-house dataset.} \label{tab5}
\resizebox{0.75\textwidth}{!}{\begin{tabular}{c|c|c|c|c}
\hline
Method/ Number of Shots & n=2 & n=4 & n=8 & n=16 \\
\hline
Logistic Regression & \textbf{59.47} & \textbf{61.9} & \textbf{65.8} &  \textbf{70.67} \\
Support Vector Machine & 57.42 & 58.64 & 61.94 & 65.28\\
Random Forests & 57.31 & 58.47 & 60.62 & 66.41 \\
MLP (end-to-end training with Stage 1) & 56.68 & 58.21 & 59.93 & 64.52 \\
\hline
\end{tabular}}
\end{table}

\section{Conclusion}
In this study, we present an efficient approach to achieve interpretable few-shot classification and zero-shot detection of retinal diseases by incorporating concept-guided prompt learning into vision-language models. Our framework provides an efficient way of creating concept bank of retinal diseases and integrating these domain knowledge into vision-language models, significantly enhancing their performance and providing added benefits of interpretability. Our research marks a significant step forward in developing highly effective and interpretable models for retinal disease diagnosis, with potential applicability to other medical image analysis domains, paving the way for real-world clinical implementation.
%
% ---- Bibliography ----
%
% BibTeX users should specify bibliography style 'splncs04'.
% References will then be sorted and formatted in the correct style.
%
% BibTeX users should specify bibliography style 'splncs04'.
% References will then be sorted and formatted in the correct style.
%
\bibliographystyle{splncs04}
\bibliography{ref}

\end{document}